\documentclass[aps,pra,twocolumn,showpacs]{revtex4}
\usepackage{amsmath,amssymb}
\usepackage{graphicx,color}
\usepackage[colorlinks,citecolor=red]{hyperref}

\newcommand\half{\frac{1}{2}}
\newcommand\unit[1]{\operatorname{#1}}
\newcommand\ket[1]{\left|#1\right\rangle}

\definecolor{gray}{gray}{0.5}

\begin{document}

\title{Superconducting Qubits Coupled to Torsional Resonators}

\author{Myung-Joong Hwang}%
\affiliation{Department of Physics, Pohang University of Science and
  Technology, Pohang 790-784, Korea}

\author{Jae-Hyuk Choi}
\affiliation{Mechanical Metrology Group, Division of Physical Metrology,
  KRISS, Korea}

\author{Mahn-Soo Choi}
\email{choims@korea.ac.kr}
\affiliation{Department of Physics, Korea University, Seoul 136-713, Korea}
\affiliation{Asia Pacific Center for Theoretical Physics, Pohang 790-784,
  Korea}
\affiliation{Department of Physics, Pohang University of Science and
  Technology, Pohang 790-784, Korea}

\date{\today}

\begin{abstract}
We propose a scheme of strong and tunable coupling between a superconducting
phase qubit and nanomechanical torsional resonator.  In our scheme
the torsional resonator directly modulates the largest energy scale (the
Josephson coupling energy) of the phase qubit, and the coupling strength is
very large.  We analyze the quantum correlation effects in the torsional
resonator as a result of the strong coupling to the phase qubit.
\end{abstract}


\maketitle

\section{Introduction}

Probing quantum mechanical properties of macroscopic objects is believed to be
a key to understand the border between the classical and quantum
physics. Nanomechanical resonators, which have high frequency of gigahertz and
low dissipation, provide a tangible system to study such macroscopic quantum
phenomena\cite{Huang03a,Knobel03a,LaHaye04a}. Coupling the resonator to the
superconducting qubits has attracted great theoretical interest as it provides
a way of control and detect the quantum behavior of the
resonator\cite{Armour02a,Irish03a,Martin04a,Tian05b,Buks06a,Wei06a,Jacobs07a,Hauss08a,Utami08a}
and a prototypical experiment has recently been
demonstrated.\cite{LaHaye09a,Oconnell10a}

Besides the fundamental aspect of the system, a nanomechanical
resonator prepared in a squeezed state can improve its noise
properties, upon which the limit of force detection sensitivity is
based, beyond the standard quantum limit.\cite{Rabl04a} An
architecture for a scalable quantum computation has also been
suggested based on the integration of the nanomechanical resonators
with the superconducting phase qubits.\cite{Cleland04a, Geller05a}




In this paper, we propose a scheme of strong and tunable coupling between a
superconducting phase qubit and nanomechanical torsional resonator.
In our scheme,
the direct modulation of the largest energy scale of the phase qubit enables a
large coupling strength.  This distinguishes our scheme from other previously
proposed schemes.  For example, in Ref.~\cite{Zhou06a}, the flexural
vibrational modes were coupled to a charge qubit by modulating the Josephson
energy, which in their case is one of the smallest energy scale of the qubit
system.
We analyze the quantum correlation effects in the torsional resonator and also
provide the noise analysis, which shows that our scheme is feasible
experimentally at the level of present technology.

The rest of the paper is organized as follows: In Section~\ref{Paper::sec:1}
we summarize the basic operational mechanism of the phase qubit and the
characteristics of the torsional resonator.  In Section~\ref{Paper::sec:2} we
analyze the coupling mechanism between the phase qubit and the torsional
resonator.  The reduced coupling constant is expressed in terms of the control
parameters of the phase qubit and the torsional resonator.  In
Section~\ref{Paper::sec:3} we discuss the possible quantum correlations
effects, especially, the squeezing of the torsional vibration mode, in the
strong coupling limit.  In Section~\ref{Paper::sec:4} we provide a detailed
noise analyses in a possible experimental realization of the scheme.  Finally
the paper is concluded in Section~\ref{Paper::sec:5}.

\begin{figure}
\centering
\includegraphics[width=7cm]{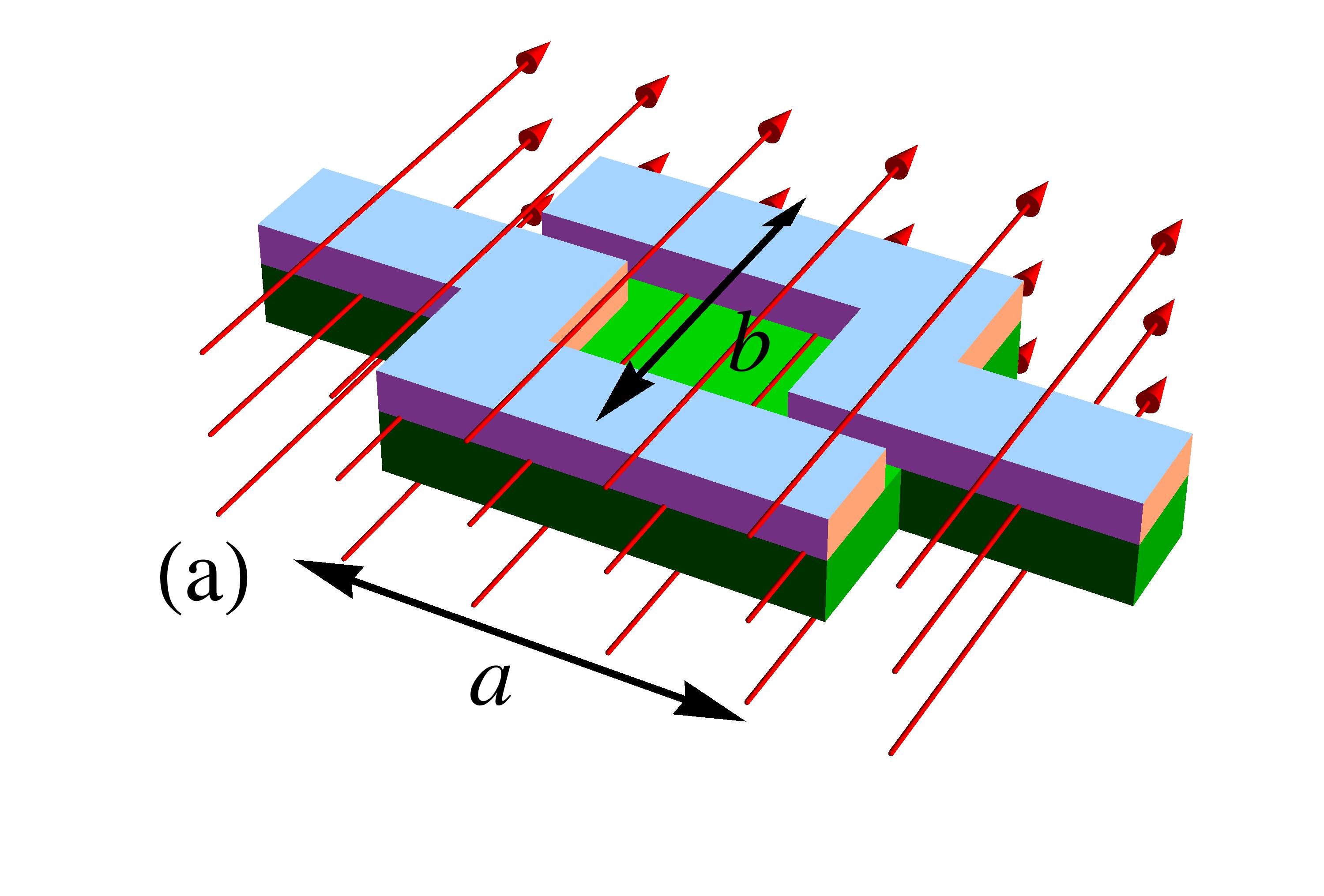}
\includegraphics[width=6cm]{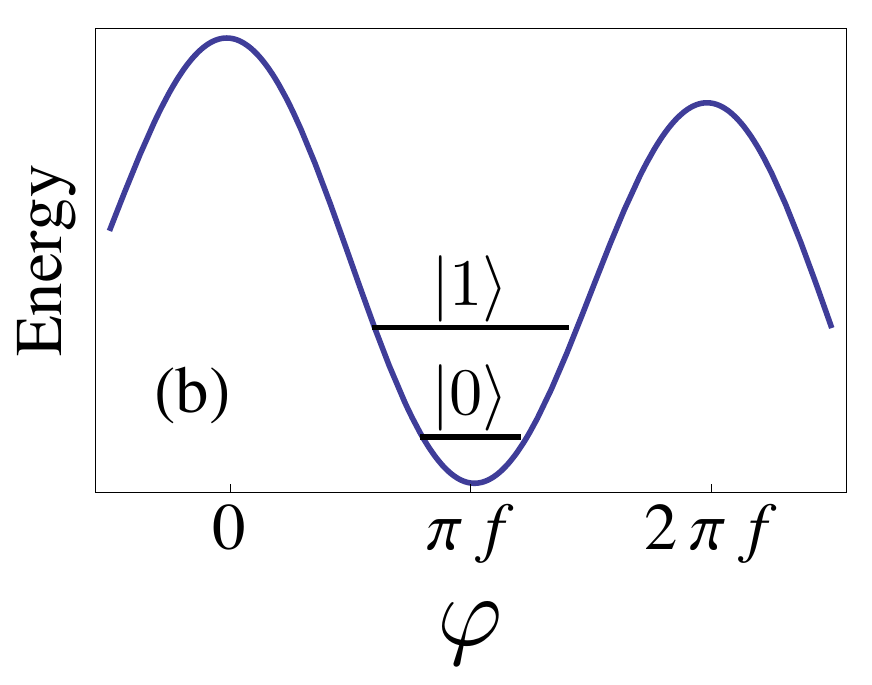}
\caption{(color on-line) (a) A schematic of superconducting phase qubit
  coupled to torsional resonator. The arrows denote the external magnetic
  field. (b) A schematic of the energy levels corresponding to the logical
  basis states $\ket{0}$ and $\ket{1}$ of a phase qubit.}
\label{Paper::fig:1}
\end{figure}

\section{Qubit and Resonator}
\label{Paper::sec:1}

A superconducting phase qubit consists of a double Josephson junction
(Fig.~\ref{Paper::fig:1}) of small size, and is described by the Hamiltonian
of the form
\begin{equation}
\label{Paper::eq:1}
H_\mathrm{qubit} = E_C n^2 - 2E_J\cos(\pi{f})\cos(\varphi-\pi{f}) \,,
\end{equation}
where the number $n$ of Cooper pairs that has tunneled through the double
junction and the phase difference $\varphi$ across the junction are quantum
mechanical conjugate variables, i.e., $[\varphi,n]=i$.  Here
$E_C=(2e)^2/2C\sim10\unit{neV}$ is the charging energy of the double junction
with total capacitance $C$, $E_J\sim50\unit{meV}$ is the Josephson coupling
energy of each junction,\cite{Endnote:1}
$f$ is the external flux (in units of the flux quantum $\Phi_0=h/2e$)
threading the loop.  In Eq.~(\ref{Paper::eq:1}), the effective Josephson
coupling
\begin{math}
E_J^\mathrm{eff}=2E_J\cos(\pi{f})
\end{math}
of a phase qubit is controlled by the external flux $f$.  A phase qubit
typically operates in the range where $k_BT\ll E_C\ll E_J$, and uses as its
computation basis the two lowest-energy states $\ket{0}$ and $\ket{1}$
confined in the potential well around $\varphi=\pi{f}$; see
Fig.~\ref{Paper::fig:1}(b).  Within the subspace spanned by the computational
basis, the qubit Hamiltonian~(\ref{Paper::eq:1}) can be written as
\begin{equation}
\label{Paper::eq:2}
H_\mathrm{qubit} \approx -\half\Omega\sigma_z
\end{equation}
where $\sigma_x,\sigma_y,\sigma_z$ are the Pauli matrices.
The level splitting $\Omega$ can be estimated by
\begin{math}
\Omega \approx \sqrt{2E_CE_J^\mathrm{eff}}
\sim 40\unit{\mu eV}
\sim 2\pi\times 10\unit{GHz}.
\end{math}\cite{Endnote:2,Blanch72a}

The torsional vibration mode of the substrate is described by a harmonic
oscillator Hamiltonian
\begin{equation}
H_\mathrm{osc} = \frac{P_\theta^2}{2I} + \half I\omega_0^2\theta^2
\end{equation}
where $P_\theta$ is the (angular) momentum conjugate to $\theta$, $I\sim
10^{-28}$--$10^{-32}\unit{kg{\cdot}m^2}$ is the rotational moment of inertia
of the torsional resonator, and $\omega_0/2\pi\sim8\text{--}800\unit{MHz}$ is
the vibration frequency.  The fluctuations of the angle $\theta$ can be
characterized by the parameter $\theta_0\equiv\sqrt{\hbar/I\omega_0}$, which
is the fluctuation in the ground state.  In typical experimental situations
$\theta_0\sim10^{-6}\text{--}10^{-7}\unit{rad}$ depending on the values of $I$
and $\omega_0$.

\section{Spin-Resonator Coupling}
\label{Paper::sec:2}

With the qubit put on the torsional resonator as in Fig.~\ref{Paper::fig:1},
the effective flux $f$ in the qubit Hamiltonian~(\ref{Paper::eq:1}) is
modulated as
\begin{math}
f = f_0\sin\theta,
\end{math}
where $\theta$ is measured relative to the direction of the external magnetic
field, and hence the qubit is coupled to the torsional vibration mode.
We point out a key advantage of this qubit-resonator coupling scheme:
As mentioned above, the phase qubit operates in the regime, where $E_J$ is the
largest energy scale.  The torsional vibration directly modulates this largest
energy scale.  This means that the coupling between the qubit and the
torsional vibrational mode can be large as demonstrated below.

\begin{figure}
\centering
\includegraphics*[width=7cm]{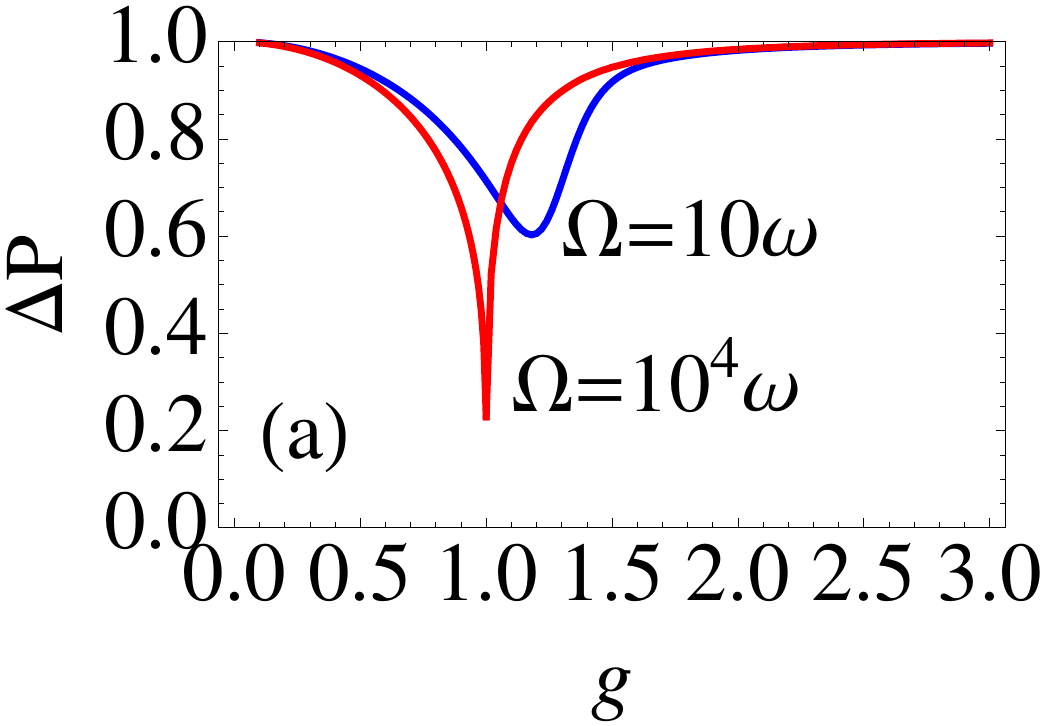}
\includegraphics*[width=7cm]{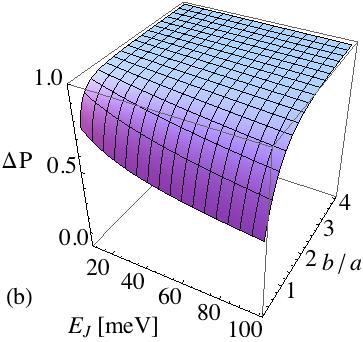}
\caption{(a) Squeezing of the torsional resonator as a function of the reduced
  coupling strength $g$ for $\Omega=10\omega$ and $\Omega=10^4\omega$. (b)
  Squeezing of the torsional resonator as a function of the Josephson energy
  $E_J$ and the ratio $b/a$ of the lateral sizes of the phase qubit.}
\label{Paper::fig:2}
\end{figure}

If we apply the external magnetic field parallel to the phase qubit plane,
then the flux modulation is given by
\begin{math}
f = f_0(\theta-\theta_e).
\end{math}
Here $f_0$ is the maximum magnetic flux (i.e., the value when the field is due
perpendicular to the qubit plane) and $\theta_e$ is the angle at equilibrium
measured from the direction of external field.  We assume that $\theta_e=0$
(non-zero $\theta_e$ slightly decreases the coupling strength by factor
$\sin\theta_e$).
Within the two-level approximation, the total Hamiltonian is given by
\begin{equation}
\label{Paper::eq:5}
H = -\half\Omega\sigma_z + \half{g}\sqrt{\Omega\omega}(a+a^\dag)\sigma_x
+ \omega a^\dag a \,.
\end{equation}
where $g$ is the dimensionless \emph{reduced coupling constant} between the
phase qubit and the torsional resonator.  Note that the oscillator frequency
has been slightly renormalized from $\omega_0$ to
\begin{equation}
\omega \equiv \omega_0\sqrt{1+2(\pi f_0)^2E_J/I\omega_0^2}
\end{equation}
due to the coupling to the phase qubit.
The renormalization of the frequency $\omega_0\to\omega$ also renormalizes the
quantum fluctuation angle $\theta_0$ to
\begin{equation}
\theta_1\equiv\sqrt{\frac{\hbar}{I\omega}} \,.
\end{equation}
The coupling constant $g$ in this case is given by
\begin{equation}
\label{Paper::eq:7}
g = \pi{f_0}\sqrt{\frac{2E_J}{I\omega^2}}
\end{equation}

The effective Hamiltonian~(\ref{Paper::eq:5}) is the well-known cavity-QED
(quantum electrodynamics) Hamiltonian for the atom-light interaction in an
optical cavity.  For optical cavities, the two-level system (or ``spin'') is
at resonance with the oscillator ($\Omega\sim\omega$), and the \emph{coupling
  energy} $g\sqrt{\Omega\omega}$ is $10^{-6}$ times smaller, at best,
than $\omega$.  In such cases, it is customary to make a so-called
rotating-wave approximation (RWA), which leads to the Jaynes-Cummings
model~\cite{Jaynes63a},
\begin{equation}
\label{Paper::eq:6}
H \approx -\half\Omega\sigma_z
+ \half g\sqrt{\Omega\omega}(a\sigma_++a^\dag\sigma_-)
+ \omega a^\dag a
\end{equation}
where $\sigma_\pm=(\sigma_x\pm i\sigma_y)/2$. The ground state of the
Jaynes-Cumming model~(\ref{Paper::eq:6}) does not exhibit any quantum
correlation effects of particular interest.  As the coupling energy
$g\sqrt{\Omega\omega}$ increases ($g\gtrsim10^{-3}$), however, the RWA breaks
down, and the ground states start to exhibit strong quantum correlation
effects such as squeezing of the oscillation mode (Fig.~\ref{Paper::fig:2}),
as discussed below.

\section{Strong coupling limits}
\label{Paper::sec:3}

Before we discuss the ``strong'' coupling limit of the qubit-resonator
composite system, we need to distinguish the limit from the
\emph{conventional} strong coupling limit.  The effective qubit-resonator
Hamiltonian (\ref{Paper::eq:5}) is an example of a more general class of
spin-boson models, which are commonly achieved in optical cavities.
Conventionally, for optical cavities, the strong coupling limit means the
coupling constant larger than the energy dissipation rate $\gamma$, so that
the coherent interaction between the two-level system and the oscillator can
be maintained.  In our case, we push the limit even further and mainly concern
about the regime, where the ground state of the qubit-resonator composite
system exhibits non-trivial quantum correlation effects.  In this paper, we
will use the squeezing in the vibrational mode as the measure of the
non-trivial quantum correlation effects.


Despite its simple form of the Hamiltonian~(\ref{Paper::eq:5}) the spin-boson
model has turned out to be highly non-trivial beyond the Jaynes-Cumming or RWA
regime~\cite{Irish07a}.
In particular, the spin-boson model~(\ref{Paper::eq:5}) is known to have a
strong squeezing effect in its ground state when the coupling energy
$g\sqrt{\Omega\omega}$ is comparable to the geometric mean
$\sqrt{\Omega\omega}$ of the two energy scales $\Omega$ and $\omega$; see
Fig.~\ref{Paper::fig:2} (a).
The manipulation of squeezed optical modes using the light-atom interaction in
an optical cavity is by now standard~\cite[see, e.g.,][]{Scully97a}.  However,
an important difference is that in our scheme the squeezing is achieved in the
\emph{static} ground state of the system.  To the contrary, in optical
cavities it can be created only by \emph{dynamical} procedures due to the weak
coupling.  That is, the two-level atoms should be prepared in a special
quantum state by means of a sequence of optical pulses before they interact
with the cavity modes and the atom-cavity interaction time should be tune
precisely depending on the initial state of the atoms.  The resulting
squeezing is thus much more difficult and less stable than in our scheme.
Another important difference is that in conventional optical cavity
$\Omega\sim\omega$ whereas in our scheme the detuning is very large
($\Omega/\omega\sim 10^4$; recall that the actual coupling energy in natural
units is given by $g\sqrt{\Omega\omega}$).

The detailed analysis and discussion on the squeezing effect in the
strong-coupled spin-boson model is out of scope of this paper.
Here we merely refer the readers to recent discussions
in \cite{Ashhab10a,1006.1989},
and define the strong coupling (also called ultra-strong coupling)
limit by the condition
\begin{math}
g \sim 1 \,.
\end{math}

The coupling constant $g$ in (\ref{Paper::eq:7}) is estimated to be $g\simeq
0.5$, surpassing the strongest coupling strength achieved so far in the
previous qubit-oscillator coupling schemes, already in the macroscopic samples
of lateral size of several micrometers ($I\sim
10^{-28}\unit{kg{\cdot}m^2}$)~\cite{Gallop02a,Evoy99a,Cleland98a}.  Here we
have put $f_0\simeq 100$ assuming an external field of $B\sim 0.1\unit{T}$
(the lower critical field of Nb, for example, is as large as $150\unit{T}$
\cite{Poole00a}).
One can enhance the coupling strength even further by using
the phase qubit with the lateral size of several hundred nanometers
and by designing its geometric shape so that the loop containing the double
Josephson junction is longer along the axis, namely, $a>b$ in
Fig.~\ref{Paper::fig:1} keeping the loop area $a\times b$ the same
(for example, $a\sim 2\unit{\mu m}$ and $b\sim 0.5\unit{\mu m}$).
The squeezing effect is also more pronounced for larger values of $\Omega$,
i.e., for Josephson junctions with larger $E_J$.
These points have been demonstrated in Fig.~\ref{Paper::fig:2} (b), where we
have plotted the uncertainty $\Delta{P}$ in the momentum quadrature as a
function of the Josephson coupling energy $E_J$ and the ratio $b/a$ of the
geometric sizes of the phase qubit.
(In the circuit QED systems based on the superconducting
circuits,\cite{Schuster07a} the theoretical limit is known to be $g\sim 1$ as
well.)
The strong coupling in our scheme is possible because the vibrational mode
directly modulates the largest energy scale ($E_J$) of the phase
qubit. Similar idea has been explored in
Refs.~\cite{Buks06a,Xue07c,Etaki08a,Buks08a,Pugnetti09a} but using the
flexural vibrations of nanomechanical beams.  Note that even for the moderate
values of $Q$-factor, $Q\sim 10^3$ \cite{Evoy99a}, the condition for the
\emph{conventional} strong coupling limit ($g\sqrt{\Omega\omega}\gg\gamma$) is
easily satisfied in our case, $g\sqrt{\Omega\omega}/\gamma\sim 10^3$.

\section{Noise analysis}
\label{Paper::sec:4}

It is important to calculate and compare the angle detection limit, for
example that of optic interferometer, and the fundamental limits due to
thermal and quantum fluctuation.
The deflection angle of a torsional resonator can be detected by measuring the
displacement at the end of its wing with a fiber-optic interferometer. A
low-power($\sim 1\unit{\mu W}$) laser light from a fiber is focused by a lens
system to a micron spot on the backside of the resonator, and reflected to
couple back into the fiber, giving a displacement-sensitive signal. For the
following analysis, we consider a silicon resonator, coupled to a
superconducting qubit, consisting of a $2\times2\unit{\mu m^2}$ rectangular
paddle suspended by narrow beams on both edges, which are $2\unit{\mu m}$ long
with a $(100\unit{nm})^2$ cross-section. To minimize heating effect due to the
probing light, it may be necessary to deposit a high-purity silver thin film
on the backside, which will play roles of a excellent reflector and a heat
sink at a low temperature of 20 mK. From our estimation, 50 nm-thick silver
coating may reduce the temperature difference from a cryogenic bath down to 5
mK, while increasing a torsional spring constant by 7 \%.
Recently, a Fabry-Perot (FP) interferometry
with a miniaturized hemi-focal cavity, developed for micrometer-sized
cantilevers, was reported to have a remarkably small noise floor as 1
$\unit{fm}/\sqrt{\unit{Hz}}$ at 1 $\unit{MHz}$ with a decreasing tail at
higher frequencies.~\cite{Hoogenboom05a} The effective noise bandwidth of the
torsional oscillator of our interest is estimated to be
$\Delta\omega=\omega/Q\sim 2\pi\times2.4\unit{kHz}$ if $Q\approx5000$,
a typical value for a micrometer-sized oscillator.~\cite{Cleland98a} The angle
detection limit for our oscillator is
\begin{equation}
\Delta\theta_{\mathrm{FP}}
= 2\sqrt{\frac{S_{\mathrm{FP}}\Delta\omega}{2\pi d^2}}
\approx 5 \times 10^{-8}\unit{rad},
\end{equation}
where $d\approx2\unit{\mu{m}}$ is the lateral size of the qubit and
$\sqrt{S_{\mathrm{FP}}}\approx 1\unit{fm}/\sqrt{\unit{Hz}}$ is the noise
floor of the FP interferometry.
The detection limit is comparable or even smaller than the quantum fluctuation
angle $\theta_1\sim10^{-7}\unit{rad}$ (for $I\sim
10^{-28}\unit{kg{\cdot}m^2}$) or larger, and enough to measure the quantum
fluctuations.

The thermal fluctuation of angle originates from the thermal energy stored in
mechanical vibration energy of the torsional resonator, and can be estimated as
\begin{equation}
\theta_T=\sqrt{k_B T/I \omega^2} \,.
\end{equation}
At an experimentally accessible low
temperature of $20\unit{mK}$ and with $I\sim 10^{-28}\unit{kg{\cdot}m^2}$, for
instance, the torsional resonator is predicted to vibrate up to
$\theta_T\approx 6.2 \times 10^{-7}\unit{rad}$.
Therefore,
the thermal fluctuation would be the main limitation to detecting the quantum
vibration.
The ratio of the thermal to quantum fluctuation is only $\sim 7$ at
$20\unit{mK}$, and can be improved even further by lowering cryogenic
temperature or by using optical or microwave cooling
technique~\cite{Rocheleau10a}. The quantum temperature $T_Q = \hbar \omega
/k_B$ is a border where the torsional resonator enters the quantum regime, and
yields $0.37\unit{mK}$. So, the thermal occupation factor $N = T/T_Q$ is $\sim
60$ when the torsional resonator is at a temperature of $T = 20\unit{mK}$,
which can be also found from an experimentally observed fluctuation angle
$\theta_T$ by $T=I\omega^2\theta_T^2/k_B$.

\section{Conclusion}
\label{Paper::sec:5}

We have proposed a scheme of strong and tunable coupling between a
superconducting phase qubit and the nanomechanical torsional resonator.  The
torsional resonator directly modulates the largest energy scale (the Josephson
coupling energy) of the phase qubit, and the coupling strength is achievable.
We have analyzed the quantum correlation effects in the torsional resonator as
a result of the strong coupling to the phase qubit.  We have also provided the
noise analysis, which shows that our scheme is feasible experimentally at the
level of present technology.

\begin{acknowledgments}
M.-S.C was supported by the NRF grant 2009-0080453 (from MEST Korea), the
KRISS grant R090-1811 (from KRISS Korea), and the Second Brain Korea 21
Program.
\end{acknowledgments}

\bibliographystyle{physrev}
\bibliography{Paper}

\begin{thebibliography}{10}
\bibitem{Endnote:1}
We assume identical junctions for simplicity, but the following discussions
  remain valid for general case.
\bibitem{Endnote:2}
A more precise value of $\Omega$ can be obtained by solving the Schr\"odinger
  equation, which is known as the Mathieu equation.
\end{thebibliography}

\end{document}